\documentstyle{article}\begin{document}
\title{A  relation between diffusion, temperature  and the cosmological constant  }
\author{ Z. Haba\\
Institute of Theoretical Physics, University of Wroclaw,\\ 50-204
Wroclaw, Plac Maxa Borna 9, Poland\\
email:zhab@ift.uni.wroc.pl\\Pacs:98.80.-k;98.80.Es;95.35.+d}\maketitle
\begin{abstract}
We show that the temperature of a diffusing fluid with the
diffusion constant $\kappa^{2}$ in an expanding universe
approaches a constant limit $T_{\infty}=\frac{\kappa^{2}}{H}$ in
its final de Sitter stage characterized by the horizon
$\frac{1}{H}$ determined by the Hubble constant. If de Sitter
surface temperature in the final equilibrium state coincides with
the fluid temperature then the cosmological constant
$\Lambda=3H^{2}=6\pi \kappa^{2}$.
   \end{abstract}

\section{Introduction}
Since the accelerated expansion has been discovered
\cite{exp1}\cite{exp2} the $\Lambda$CDM model  became the standard
model of cosmology. There are many aspects of this model which
still need exploration, e.g., the relation between the dark matter
(DM) and  dark energy (DE) densities (coincidence). Such properties can be
explained in models with some interactions between universe
constituents. In \cite{hss} we have suggested a model in which the
DM energy-momentum non-conservation is a consequence of  its
diffusive energy gain from an environment described by an ideal
fluid (as a model of dark energy). In such a scheme equations of
the relativistic diffusion have as a solution
 a phase space distribution in the form of the J\"uttner
distribution \cite{jut} with a time-dependent temperature. The
scale factor can increase exponentially for a large time  in
Einstein gravity of such a diffusive fluid \cite{habacqg} in
agreement with the $\Lambda$CDM prediction. In a complete theory
baryons and radiation will be described as excitations of quantum
fields which are to live in a space-time approaching at large time
de Sitter space-time. It is known that if quantum fields at finite
temperature $T$ are to exist in de Sitter space then there must be
a relation between the temperature and the de Sitter horizon
\cite{figari}\cite{fullingrep}. A simple reason is that at an
imaginary time the model of de Sitter space as a pseudo-sphere
becomes the sphere. The finite temperature quantum fields to be
periodic in the imaginary time (KMS condition) must be periodic on
the sphere. Hence, $\beta=\frac{1}{T}= 2\pi r$ where
$r=\frac{1}{H}$ is the radius of the sphere. This  relation can be
considered as an extension of the black hole horizon-temperature
correspondence \cite{bekenstein}\cite{hawking}. In de Sitter case
the relation has been confirmed by  Gibbons and Hawking
\cite{gibbons} through a calculation of the temperature of
particles created from the event horizon. In this paper we assume
that DM consists of relativistic diffusing particles ( see
\cite{hss} \cite{cal1}\cite{cal2}). The relativistic diffusion can
be defined in a coordinate independent way as a diffusion on the
tangent bundle \cite{franchi}. We restrict ourselves to the study
of the relativistic diffusion
 in a homogenous space-time. We show that solutions of the
diffusion equation approach an equilibrium limit which depends on
the diffusion constant and the Hubble constant. Assuming that DM
and baryonic matter (described by quantum fields) coexist in de
Sitter space we obtain the relation $H^{2}=2\pi \kappa^{2}$
between the diffusion constant $\kappa^{2}$ and the Hubble
constant $H$. Finally, we point out that if at large time the
diffusion becomes non-relativistic then our conclusion concerning
the relation between the diffusion constant and the cosmological
constant does not change.
\section{Diffusive DM-DE interaction}
We consider an energy-momentum tensor $T^{\mu\nu}$ consisting of a
separately  conserved baryonic part and $T_{D}^{\mu\nu}$
describing the dark sector. In order to approach the coincidence
problem we
 assume an interaction between DM and DE. If $T_{D}^{\mu\nu}=T_{M}^{\mu\nu}+T_{E}^{\mu\nu}$
is to be conserved then we have a relation between the
non-conservation laws of DE and DM
\begin{equation}
-\nabla_{\mu}T_{E}^{\mu\nu} =\nabla_{\mu}T_{M}^{\mu\nu}\equiv
3\kappa^{2}J^{\nu},
\end{equation} where $T_{M}^{\mu\nu}$
is the energy-momentum
 of DM and $T_{E}^{\mu\nu}$ corresponds to DE. The rhs of eq.(1)
could be treated   as a definition of $J^{\nu}$, $\kappa^{2}$ is a
parameter
 which measures the strength of the DM-DE interaction.
 Our crucial assumption is that the four-vector $J^{\nu}$
 appearing on the rhs of eq.(1) is conserved
 \begin{equation}
\nabla_{\mu}J^{\mu}=0.
\end{equation}
Such a conserved current can describe a flow of particles with the
density $J^{0}$. If additionally we assume that the particle
current is realized as a stream of particles with the phase space
distribution $\Omega(p,x)$ then the current should be expressed as
\begin{equation}
J^{\mu}=\sqrt{g}\int \frac{d{\bf p}}{p^{0}}p^{\mu}\Omega.
\end{equation}
The  energy-momentum of a stream of particles is determined by the formula
\begin{equation}
T_{M}^{\mu\nu}=\sqrt{g}\int \frac{d{\bf
p}}{p^{0}}p^{\mu}p^{\nu}\Omega.
\end{equation}In eqs.(3)-(4) $g_{\mu\nu}$ is the Riemannian metric,
the momenta satisfy the mass-shell condition
$g_{\mu\nu}p^{\mu}p^{\nu}=m^{2}$ (the velocity of light $c=1$) and
$g=\vert \det g_{\mu\nu}\vert$. The definitions (3)-(4) come from
relativistic dynamics. $J^{\nu}$ and $T_{M}^{\mu\nu}$ are conserved as
a consequence of the Liouville equation if the number of particles
is preserved and there is no exchange of energy with the
environment,i.e., $\kappa^{2}=0$. We wish to find a realization of
the conservation laws (1)-(2). If the phase space distribution
$\Omega$ is to satisfy a differential equation which is to reduce
to the Liouville equation when $\kappa^{2}=0$ then
the modified differential equation must be a diffusion equation. The diffusion is determined
 in the unique way by the requirement that the diffusing
particle moves on the mass-shell (see
\cite{dudley}\cite{franchi}\cite{habapre}\cite{calo}).

Let $p^{j}$ be spatial coordinates  on the mass-shell
$p^{2}=m^{2}$ . We define the Riemannian metric on the mass-shell
as the one induced from the pseudo-Riemannian metric $g_{\mu\nu}$
on the space-time
\begin{displaymath}
ds^{2}=g_{\mu\nu}dp^{\mu}dp^{\nu}=-m^{2}G_{jk}dp^{j}dp^{k},
\end{displaymath}
where $p^{0}$ is expressed by $p^{j}$.

The inverse matrix is  (we assumed that $ g_{0k}=0$)
\begin{equation}
G^{jk}=-g^{jk}m^{2}+p^{j}p^{k}.
\end{equation}
Next,
\begin{displaymath}
G\equiv \det(G_{jk})=m^{-4}\det(g_{jk})\omega^{-2},
\end{displaymath}
where
\begin{displaymath}
\omega^{2}=m^{2}-g_{jk}p^{j}p^{k}.
\end{displaymath}
 We define   diffusion as a stochastic process  generated by
the Laplace-Beltrami operator $\triangle_{H}^{m}$ on
 the mass-shell
\begin{equation}
\triangle_{H}^{m}=\frac{1}{\sqrt{G}}\partial_{j}G^{jk}\sqrt{G}\partial_{k},
\end{equation}
 where
 $\partial_{j}=\frac{\partial}{\partial
p^{ j}}$.

The transport equation for the  diffusion generated by
$\triangle_{H}$ reads
\begin{equation}
\begin{array}{l}
(p^{\mu}\partial^{x}_{\mu}-\Gamma^{k}_{\mu\nu}p^{\mu}p^{\nu}\partial_{k})\Omega=
\kappa^{2}\triangle^{m}_{H}\Omega,
\end{array}\end{equation} where $\kappa^{2}$ is the diffusion constant
and $\partial_{\mu}^{x}=\frac{\partial}{\partial x^{\mu}}$. Both
operators:  the infinitesimal transport  on the lhs of eq.(7) as
well as the generator of diffusion on the rhs preserve the
mass-shell. The operator $-\triangle^{m}_{H}$ is self-adjoint and
non-negative in the Hilbert space $L^{2}(\frac{d{\bf p}}{p^{0}})$.
Eq.(7) does not depend on the choice of coordinates on the tangent
bundle \cite{franchi}. It can be checked by direct calculations
using eq.(7) that $\nabla_{\mu}T^{\mu\nu}_{M}=3\kappa^{2}J^{\nu}$.
For the first part of the equality (1) we need a model of the dark
energy. In \cite{hss} we have chosen an ideal fluid as a model of
$T^{\mu\nu}_{E}$. We have determined its energy density
$\rho_{de}$ from eq.(1). The pressure $p_{de}$ is defined as
$p_{de}=w\rho_{de}$ where $w$ is a free parameter. The Einstein
equations
\begin{displaymath}
R_{\mu\nu}-\frac{1}{2}g_{\mu\nu}R=\frac{1}{3}(T_{M}^{\mu\nu}+T_{E}^{\mu\nu})
\end{displaymath}(where $R_{\mu\nu}$ is the Ricci tensor)
have asymptotic exponentially growing solutions if $w=-1$.
\section{Solutions of the diffusion equation}
We restrict our discussion to the metric \begin{equation}
ds^{2}=dt^{2}-a^{2}d{\bf x}^{2}.
\end{equation}
We change variables
\begin{displaymath}
{\bf p}=a^{-1}{\bf q}.
\end{displaymath}
Let
\begin{equation}
\omega_{1}=\sqrt{m^{2}+{\bf q}^{2}}.
\end{equation}
Then, the diffusion equation (7)(neglecting the spatial
dependence) reads
\begin{equation}
\begin{array}{l}
(\partial_{t}-Hq^{k}\partial_{k})\Omega=\kappa^{2}
\partial_{k}\Big(\omega_{1}^{-1}(m^{2}\delta^{jk}+q^{j}q^{k})\partial_{j}\Big)\Omega,
\end{array}\end{equation}
 where $H=a^{-1}\partial_{t}a$ and
 $\partial_{j}=\frac{\partial}{\partial
q^{ j}}$.

We cannot solve eq.(10) explicitly in general. However, if $H=const
$ ,i.e. $a=\exp(Ht)$, then we obtain a particular solution (we
have got this solution in \cite{habacqg0} using slightly different
notation)
\begin{equation}
\Omega^{deS}(q)=\exp(-3Ht)\exp(-\frac{H\omega_{1}(q)}{\kappa^{2}}).
\end{equation}
The time-dependent factor in eq.(11) cancels if we calculate
expectation values of observables ${\cal O}$ with respect to the
normalized distribution
\begin{equation}
\langle {\cal O}\rangle_{\Omega}=\Big(\int d{\bf
q}\Omega\Big)^{-1}\int \frac{d{\bf q}}{\omega_{1}}\Omega
\omega_{1}{\cal O}\equiv
(\Omega,\omega_{1})^{-1}(\Omega,\omega_{1}{\cal O}),
\end{equation}
where on rhs of eq.(12) we introduce a scalar product in the
Hilbert space $L^{2}(\frac{d{\bf q}}{\omega_{1}})$ of functions
square integrable with respect to the Lorentz invariant measure on
the mass shell (in this Hilbert space the diffusion generator is
self-adjoint).

The solution (11) describes the J\"uttner equilibrium distribution
\cite{jut} corresponding to the temperature (we use the units with
the Boltzmann constant $k_{B}=1$ and $\hbar=1$)
\begin{equation}
T_{\infty}=\frac{\kappa^{2}}{H}.
\end{equation}
The J\"uttner distribution is the canonical distribution
determined by the maximum entropy principle (assuming the fixed
value of the energy $\langle p^{0}\rangle$). The entropy is
defined by the integral (in terms of the original momenta ${\bf
p}$ of sec.2)
\begin{equation}
S=-\int d{\bf x}g d{\bf p}\Omega\ln\Omega
\end{equation}
In \cite{habacqg} we have considered the time evolution of the
J\"uttner distribution  (we choose $ a(t_{0})=1 $ and we denote
$T_{0}=T(t_{0})$)
\begin{equation}
\Omega(t=t_{0},T_{0},q)=\exp(-\frac{\vert{\bf q}\vert}{T_{0}})
\end{equation}
for an arbitrary scale factor $a$ in the limit $m=0$. We have
shown that the solution of eq.(7) with the initial condition (15) is
again the J\"uttner distribution
\begin{equation}\begin{array}{l}
\Omega^{(m=0)}(t)=T_{0}^{3}(T_{0}+\kappa^{2}A)^{-3}\exp\Big(-\frac{a}{T_{0}+\kappa^{2}A}\vert{\bf
q}\vert\Big),\end{array}
\end{equation}
where
\begin{equation}
A(t)=\int_{t_{0}}^{t}a(s) ds.\end{equation}

The temperature corresponding to the phase space distribution $\Omega(t)$ can again  be determined either from
the maximum entropy principle as the factor in front of
$\vert {\bf q}\vert$ in eq.(16) or  by calculating the entropy (14) and the energy
$E$ defined by
\begin{equation}
E=\int d{\bf x}\sqrt{g} T_{M}^{00}
\end{equation}
Then, by general rules of statistical physics
\begin{equation}
\frac{1}{T}=\frac{\partial S}{\partial E} \end{equation} The result
is
\begin{equation} T(t)=\frac{T_{0}+\kappa^{2}A}{a}.
\end{equation}
If for a large time $a(t)\simeq\exp(\lambda t)$ then
$\Omega^{(m=0)}\rightarrow
T_{0}^{3}(\frac{\lambda}{\kappa^{2}})^{3} \Omega^{deS}$(with $m=0$
and $\lambda=H$ in eq.(11)). Moreover, if both $A(t)$ and $a(t)$
tend to infinity at large $t$ and the limit of $T(t)$ in eq.(20)
is $\frac{\lambda}{\kappa^{2}}$, then, $\lim_{t\rightarrow\infty}
a(t)\exp(-\lambda t)=const\neq 0$. Hence, a finite limit of the
expectation value (12) with $\Omega^{(m=0)}$ implies the de Sitter
growth of $a$.

If instead of the J\"uttner distribution (15) we take as an
initial condition a superposition of J\"uttner distributions
($\mu\geq 0$ )

\begin{equation}\Omega(t_{0},q)=\int_{0}^{\infty}dT_{0}\mu(T_{0})
\exp( -\frac{\vert{\bf q}\vert}{T_{0}})
\end{equation}
then
\begin{equation}
\Omega(t,q)=\int_{0}^{\infty}dT_{0}\mu(T_{0})T_{0}^{3}(T_{0}+\kappa^{2}A)^{-3}\exp\Big(-\frac{a}{T_{0}+\kappa^{2}A}\vert{\bf
q}\vert\Big)\end{equation} If we allow complex $T_{0}$ then
eq.(21) represents a large class of initial conditions (including
Fourier transforms) and eq.(22) a general solution of the
diffusion equation with $m=0$.

 Assuming an exponential growth of $a$
(as discussed below eq.(20)) and using Lebesgue dominated
convergence theorem we  obtain a limiting behaviour of the
solution (22)
\begin{equation}\begin{array}{l}lim_{t\rightarrow\infty}\exp(3Ht)\Omega(t,q)\cr =\exp( -\frac{H\vert{\bf
q}\vert}{\kappa^{2}})\int_{0}^{\infty}dT_{0}\mu(T_{0})
T_{0}^{3}(\frac{H}{\kappa^{2}})^{3}\end{array}
\end{equation}
if $ \int_{0}^{\infty}dT_{0}\mu(T_{0}) T_{0}^{3}<\infty$. The
exponential factor $\exp(3Ht)$ does not contribute to the
expectation value (12). Hence, in the sense of the expectation
values the limit $t\rightarrow \infty$ of the solution of the
relativistic diffusion equation with the initial condition (21) is
again the J\"uttner distribution with the temperature
$\frac{\kappa^{2}}{H}$.

We are unable to derive an explicit solution of the diffusion
equation for arbitrary $a$ and $m>0$. We apply a perturbative
method. We consider the initial condition
\begin{equation}\Omega(t_{0},q)=\int_{0}^{\infty}dT_{0}\mu(T_{0})
\exp( -\frac{\omega_{1}(q)}{T_{0}}).
\end{equation}
We look for solutions of eq.(7) with $m\neq 0$  in the form
\begin{equation}
\Omega(t,q)=\int_{0}^{\infty}dT_{0}\mu(T_{0})L(t)\exp(-\omega_{1}
\alpha(t))f_{t}(\vert {\bf q}\vert,T_{0})
\end{equation}
 with
\begin{equation}
\alpha(t)=\frac{a}{T_{0}+\kappa^{2}A}
\end{equation}
and
\begin{equation}
L(t)=T_{0}^{3}(T_{0}+\kappa^{2}A)^{-3}.
\end{equation}
We would like to show that  similarly as in the massless case (22)
if the space-time asymptotically tends to de Sitter space then
$\Omega_{t}\rightarrow\Omega^{deS}$ (11).

Inserting eq.(25) in eq.(7) we obtain an equation for $f$
\begin{equation}\begin{array}{l}
\partial_{t}f-\kappa^{2}\omega_{1}\partial_{q}^{2}f-\kappa^{2} (\kappa^{-2}Hq-2\alpha
q+
\frac{3q}{\omega_{1}}+\frac{2m^{2}}{q\omega_{1}})\partial_{q}f=\frac{m^{2}}{\omega_{1}}\partial_{t}\alpha f,
\end{array}\end{equation}where
$\partial_{q}=\frac{\partial}{\partial q}$ and $q=\vert {\bf
q}\vert$. Let us note that if $a(t)=\exp(\lambda t)+r(t)$, where
$r(t)$ is a slowly varying function, then
\begin{displaymath}\begin{array}{l}
H=(\exp(\lambda t)+r(t))^{-1}(\lambda\exp(\lambda t)\cr
+\partial_{t}r(t))\simeq \lambda +\exp(-\lambda t)R(t),\end{array}
\end{displaymath}where $R(t)$ is a slowly varying function.
Hence, $H\simeq \lambda$ and we can treat $H$ as a constant up to
an exponentially small correction. For the same reason
$\alpha-\kappa^{-2}H=\exp(-Ht)R_{1}$ where $R_{1}$ is a slowly
varying function and $\partial_{t}\alpha=\exp(-Ht)R_{2}$ where
$R_{2}$ is a slowly varying function. The exponentially decaying
corrections will be shifted to the rhs of eq.(28). Now,
 eq.(28) can be expressed as
\begin{equation}\begin{array}{l}
\partial_{t}f+{\cal
M}f=(v(t,q)\partial_{q}+\frac{m^{2}}{\omega_{1}}\partial_{t}\alpha)f,
\end{array}\end{equation}
where
\begin{equation}\begin{array}{l}
-{\cal M}=\kappa^{2}\omega_{1}\partial_{q}^{2}+\kappa^{2}
(-\kappa^{-2}Hq+\frac{3q}{\omega_{1}}+\frac{2m^{2}}{q\omega_{1}})\partial_{q}.
\end{array}\end{equation}
${\cal M}$  depends neither  on time nor on $T_{0}$. The functions
$v(t,q)$ and $\partial_{t}\alpha$ on the rhs of eq.(29) are
decreasing as $\exp(-Ht) $ times a slowly varying function. We can
rewrite eq.(29) as an integral equation with the initial value
$f_{0}=1$ (i.e., we assume  the initial condition (24)).

\begin{equation}
f_{t}(q)=1+\int_{t_{0}}^{t}\exp(-(t-s){\cal
M})\Big(\exp(-Hs)R_{3}(s)q\partial_{q}+\frac{m^{2}}{\omega_{1}}\exp(-Hs)R_{4}(s)\Big)f_{s}ds,
\end{equation}where $R_{k}$ are slowly varying functions.
 We  solve eq.(31) by iteration. We can show that in the sense  of the convergence of expectation values (12)
  owing to the exponential decrease of the
integrand on the rhs of eq.(31) each term of the iteration series
tends to zero as $t\rightarrow \infty$. Hence, $f_{t}\rightarrow
1$.

 Let us still consider the non-relativistic limit of the diffusion equation (7). In the  limit
\begin{displaymath}
m^{2}a^{2}\rightarrow \infty
\end{displaymath}
eq.(7) reads \cite{habacqg}\cite{hss}
\begin{equation}
m^{-1}\kappa^{-2}(\partial_{t}-Hq^{j}\partial_{j})\Omega=\triangle_{\bf
q}\Omega.
\end{equation}
$\triangle_{\bf q}$ is the Laplacian and  \begin{equation} A_{NR}=
2\int_{t_{0}}^{t} ds a^{2}.
\end{equation}
A solution of the diffusion equation (32) which starts from the
Maxwell-Boltzmann distribution (the non-relativistic approximation
of the J\"uttner distribution) is \cite{habacqg}\cite{hss}
\begin{equation} \Omega_{NR}(t,{\bf q})=T_{0}^{\frac{3}{2}}
(\kappa^{2}A_{NR}+T_{0})^{-\frac{3}{2}}\exp \Big(-a^{2}\frac{{\bf
q}^{2}}{2m(\kappa^{2}A_{NR}+T_{0})}\Big).
\end{equation}
Using eq.(34) we can solve the diffusion equation with an initial
condition
\begin{equation}
\Omega(t_{0},{\bf q})=\int_{0}^{\infty}dT_{0}\mu(T_{0})\exp
\Big(-\frac{{\bf q}^{2}}{2mT_{0}}\Big).
\end{equation} The solution reads
\begin{equation}
\Omega_{NR}(t_,{\bf
q})=\int_{0}^{\infty}dT_{0}\mu(T_{0})\Omega_{NR}(t,{\bf q}).
\end{equation}
 It follows from eq.(34) and from eq.(19) that in the
non-relativistic limit the temperature is
\begin{equation}
T(t)=\frac{T(t_{0})+\kappa^{2}A_{NR}}{a^{2}}.\end{equation}If for
a large time $a(t)\simeq \exp(Ht)$ (plus a slowly varying function) then
\begin{equation}\begin{array}{l}
\exp(3Ht)\Omega_{NR}(t,{\bf q})\cr\rightarrow \exp \Big(-\frac{H
{\bf
q}^{2}}{2m\kappa^{2}}\Big)(\frac{H}{\kappa^{2}})^{\frac{3}{2}}\int_{0}^{\infty}dT_{0}\mu(T_{0})T_{0}^{\frac{3}{2}}
\end{array}\end{equation} Hence, the limiting temperature is the same as
in the relativistic case (13) $ T(t)\rightarrow
T_{\infty}=\kappa^{2}H^{-1} $ (we obtain the result (38) even if
the temperature $T_{0}$ in eqs.(34)-(35) is a  complex number ).

If we  assume that the diffusing  dark matter is in equilibrium
with the baryonic matter described by quantum fields defined on
the de Sitter space \cite{figari}\cite{gibbons} then
$T_{\infty}=T_{deS}=\frac{H}{2\pi}$. Hence, from eq.(13)
\begin{equation}
H^{2}=2\pi\kappa^{2}
\end{equation}

\section{Conclusions} We have shown for a class of initial
conditions that (in the sense of expectation values of
observables)
 solutions of the diffusion equation in an exponentially expanding universe tend to the J\"uttner
equilibrium distribution with the temperature
$T_{\infty}=\frac{\kappa^{2}}{H}$. It is widely recognized that
quantum field theory at finite temperature on de Sitter space has
the temperature related to the event horizon. In a generalized
thermodynamics
\cite{bekenstein}\cite{haw}\cite{gib}\cite{davis1}\cite{davis2}
the second law leads to the conclusion that the temperature inside
the horizon must be bigger than the horizon temperature which is
achieved when the system is approaching the equilibrium. In a
classical approximation the energy momentum of the quantum
(baryonic) fields at the temperature $T_{\infty}$ will be described by classical fluids
of temperature $T_{\infty}$. In our model the DM
energy-momentum non-conservation results from a diffusion in a
fluid of DE.
 If we accept this scheme then
the diffusion constant in an expanding universe is determined by
the Hubble constant (or vice versa). Inserting the current value
of the Hubble constant we obtain $\kappa\simeq 10^{-42} GeV$ . The small value of the
diffusion constant follows from the weak interaction between DM
and DE fluids. The weakness of this interaction could be regarded
as an answer, $\Lambda=6\pi \kappa^{2}$, to the cosmological
constant problem \cite{weinberg}.

{\bf Acknowledgement} The research is supported by the NCN grant
DEC-2013/09/BST2/03455

\end{document}